\documentclass[a4paper,12pt]{article}
\usepackage{amsmath, amsthm, amsfonts, amssymb}
\usepackage[svgnames]{xcolor}
\colorlet{color1}{NavyBlue}
\usepackage[colorlinks=true,allcolors=color1]{hyperref}
\usepackage{physics}
\usepackage{dsfont}
\usepackage{graphicx}
\usepackage{cleveref} 
\usepackage{xfrac}
\usepackage{cite}

\usepackage[margin=1in]{geometry}

\title{\vspace*{-1cm} \par \Large\bf Way down in the hole... and up again
}

\author{
	Valentin Boyanov, \\
	\small\it Departamento de F\'{i}sica Te\'{o}rica and IPARCOS, Universidad Complutense\\
	\small\it de Madrid, 28040 Madrid, Spain\\
}

\def\fecha{29/03/2022}
\date{\fecha}

\begin{document}

\maketitle

\setcounter{page}{0}

\thispagestyle{empty}
\vspace{0.5cm}
\hrule
\begin{abstract}
	\noindent
I argue that an approach which uses an appropriate admixture of both classical and semiclassical effects is essential for understanding the ultimate fate of gravitational collapse and the nature of black holes. 
I provide an example of a problem which pushes the boundaries of what is known in both the classical and semiclassical approaches: the evolution of the inner horizon of a black hole. I show that solving this problem requires considering perturbations of both classical and semiclassical origin.
In fact, it has been found that classical mass inflation might be counteracted by a semiclassical tendency for the inner horizon to inflate outward.
\end{abstract}

\hrule
\vspace{0.5cm}

\centerline{Essay written for the Gravity Research Foundation 2022 Awards for Essays on Gravitation}

\vspace{0.5cm}

\underline{E-mail}: vboyanov@ucm.es

\newpage

\section{Introduction}

Within the past 60 years, black holes have gone from being a mere mathematical curiosity in highly symmetric solutions of the Einstein field equations \cite{Thorne1970}, to being among the most actively studied astrophysical objects \cite{LIGO,EHT}. Accordingly, our theoretical understanding of these objects has undergone a vast transformation. The very first model of black hole (BH) formation by Oppenheimer and Snyder \cite{Oppenheimer1939} described the collapse of a spherically-symmetric homogeneous pressureless fluid (or dust cloud) into a Schwarzschild BH, with its simple spacelike singularity enclosed by an event horizon. By contrast, nowadays, a complete picture of a dynamically formed BH must incorporate elements like the mass inflation instability \cite{PoissonIsrael89}, the BKL singularity \cite{Belinsky1970}, semiclassical evaporation \cite{Hawking1975} and potential information loss \cite{Page1993}, among others.
Given all this complexity, it is hardly surprising that there is a tendency to over-simplify the problem in practice. In fact, our intuition about BHs is still largely based on that first model of a collapsing dust cloud, or even directly on the vacuum Schwarzschild solution itself, and tends to ignore the more intricate details.


Research into BHs has seemingly gone down two different routes, each one choosing to keep different elements of the basic Schwarzschild solution as simplifying assumptions. 
The first route can be broadly characterised by the fact that it preserves the classical notion 
of an eternal trapped region enclosed by an event horizon. It considers (at least implicitly) that anything which happens within the trapped region stays there. 
It therefore guides one to a belief that the detailed description of spacetime beyond the horizon surface is a mere theoretical curiosity, with concepts like extensions into other universes \cite{Dafermos2017} which are of little consequence to anything actually observable.
Curiously, aside from purely classical studies, this assumption is often tacitly made even in the semiclassical context, e.g. in works which analyse the divergence of the quantum vacuum energy content at the Cauchy horizon of charged and spinning BHs \cite{Birrell1978,BalbinotPoisson93,Hollands2020a,Hollands2020b,Ori2019,Zilberman2022}. 
This picture, however, contains a potential conceptual mismatch with the notion of semiclassical BH evaporation.

The second route is one in which the trapped region is considered as evaporating, but the interior it reveals is none other than the simplest of all classical backgrounds: the vacuum Schwarzschild solution \cite{Hawking1974,Stephens1993,Kaplan2018,Page2004,Giddings2019}, with modifications encountered only at the very core of the configuration, where Plackian curvatures, and consequently strong quantum effects, are expected to be present. An example of this approach can be found in the simplest explanations of the information loss problem in BHs, which indeed do not dwell on the specific form of the putative singular region~\cite{Preskill1992,Susskind2008,Giddings2019}.
In defence of this view, it is true that at times it has been argued that the classical evolution of BHs essentially reduces to the formation of a Schwarzschild-like singularity even when considering a more complete picture involving mass inflation \cite{Poisson1990}. However, a mass inflation interior at finite times actually appears to possess quite a different classical structure \cite{Marolf2012}.

The dissonance between the research conducted within these two routes, despite the fact that they analyse the exact same physical system, is an interesting phenomenon, an explanation of which would likely take us beyond the realm of physics, which, of course, in not the purpose of the present work. Our goal is rather to argue why---and give some indications as to how \cite{Barcelo2022}---these two routes should converge into a single, holistic semiclassical description of BHs. 
Of course, this may still not constitute a complete description of BHs, as there might be regions not analysable in terms of just semiclassical physics~\cite{Ashtekar2005,Mathur2005}. However, it would be a major step forward in this direction.

\section{A unifying example: semiclassical mass inflation}

The semiclassical Einstein equations consist of taking the vacuum expectation value of the renormalised stress-energy tensor operator (RSET), constructed from a quantum field, as part of the source of curvature,
\begin{equation}\label{semicl}
	G_{\mu\nu}=8\pi G \left(T_{\mu\nu}^{\rm \,class}+\expval{T_{\mu\nu}}\right).
\end{equation}
Here $T_{\mu\nu}^{\rm \,class}$ represents the effectively classical matter content, and $\expval{T_{\mu\nu}}$ is the RSET. The latter is suppressed by a Planck constant, and is thus generally small when compared to its classical counterpart.

Therefore, places where semiclassical corrections to geometries can become relevant are, in some sense, special. Since the RSET contains terms which depend on the spacetime curvature, one such place is regions where curvature is close to becoming Planckian. 
Although the validity of the semiclassical approximation itself (considered as an expansion in $\hbar$) is difficult to maintain in such scenarios, at a heuristic level some interesting results can still be obtained in this regime \cite{Anderson1983}.
Another special place is determined to be the outer horizon of BHs, particularly in the absence of any dynamics-inducing classical matter. Then, even if the RSET is small, being the only source of dynamics, it can become quite relevant: this is the regime considered when analysing Hawking evaporation \cite{Hawking1974}.

\begin{figure}
	\centering
	\includegraphics[scale=1.2]{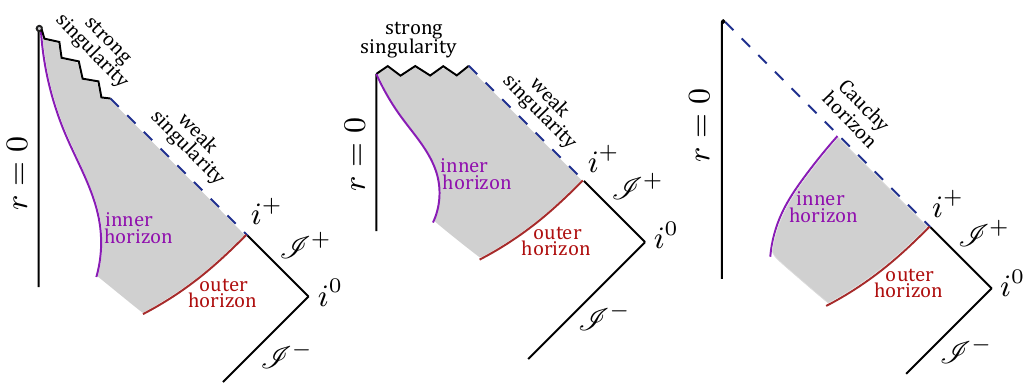}
	\caption{Causal structure of the future region of each of the three possibilities for classical mass inflation. Left: the inner horizon tends to the origin asymptotically in $v$, leading to a strong component in the null singularity. Middle: the inner horizon reaches the origin at finite $v$, forming a Schwarzschild-like singularity. Right: the inner horizon stabilises at a finite radius and mass inflation is halted.}
	\label{f1}
\end{figure}

Yet another special place turns out to be the vicinity of the \textit{inner horizon} of BHs. Arguments regarding this go back over 40 years \cite{Birrell1978}. Actually, rather than pertaining to the inner horizon at finite times (which exists only in dynamically formed BHs, and not in analytical extensions of stationary solutions; see fig.~\ref{f1}), these arguments refer to the Cauchy horizon located at infinite coordinate time $v$ (the Eddington-Finkelstein advanced time). However, much like how the classical mass inflation instability related to the Cauchy horizon has a dramatic effect on the interior of the BH at finite times \cite{Marolf2012}, it is not a big leap to expect that something equally dramatic may happen semiclassically, especially given the fact that the RSET seems to become the dominant source of dynamics when the formation of a Cauchy horizon is approached (as argued in \cite{Hollands2020a,Hollands2020b,Ori2019,Zilberman2022}, where a case is made for the presence of a RSET-sourced strong singularity).

Therefore, the vicinity of a BH inner horizon, being prone to large backreaction from both classical and semiclassical perturbations, seems like an ideal place to attempt a unified description containing both. In a recent work \cite{Barcelo2022}, the present author and collaborators made just such an attempt using a simple spherically symmetric model. Now, I will present the context for this work, as well as sketch out its main conclusions.
We will first look at classical and semiclassical backreaction at the inner horizon individually, and then see what may come of bringing them together.

\subsection{Classical inner horizon evolution}

Generic BH solutions have an inner horizon, be it due to the presence of angular momentum, electric charge, or an effectively classical regular core~\cite{Wald1984,Hayward,Ansoldi}. When a BH is formed dynamically, this inner horizon is not directly identified with a Cauchy horizon (as is often the case when working with analytical extensions). Rather, the inner horizon is just the lower boundary of a trapped region, as shown in fig.~\ref{f1}. Geodesics which approach the inner horizon (or in the case of a mass inflation interior, which approach points of its past trajectory) may end up incomplete and extendable in the $v\to\infty$ limit. Then, a Cauchy horizon would indeed be generated.

The presence of a long-lived inner horizon leads to the well known mass inflation instability \cite{PoissonIsrael89,Ori1991,Page1991,Barrabes1990,Brady1992,Brady1995,Ori1992,Ori1998}. Backreaction from classical perturbations from the exterior (which, for asymptotically flat configurations, decay as an inverse polynomial in time) can trigger an exponential (or at least polynomial \cite{Raul2021}) growth of the Misner-Sharp mass in the interior of the BH. In our recent work \cite{Barcelo2022}, we used a spherically symmetric BH model in which the perturbations were represented by ingoing and outgoing null shells. Using this simple setup, we analysed the possible outcomes of mass inflation for generic spherical BHs with an inner horizon.

Interestingly, we found that the instability is only triggered when the core of the BH has a particularly malleable nature under mass perturbations. In particular, mass inflation only occurs generically when the radial position of the inner horizon $R_{\rm i}$ is related as an inverse polynomial to the total mass $M$,
\begin{equation}\label{ingredient1}
R_{\rm i}\simeq\frac{b}{M^p},
\end{equation}
where $b$ and $p$ are positive. Note that this relation implies that $R_{\rm i}$ approaches zero when the mass is very large. Although this is indeed the case for charged and rotating BHs (when their charge and angular momentum are kept fixed), one can imagine that a dynamical generalisation of regular BH models \cite{Hayward,Ansoldi} should behave quite differently, given that the reason for their construction is the assumption that a regular core is formed when matter is compressed sufficiently (e.g. due to an effective change in its equation of state leading to negative pressures).

For the inner horizon, the outcomes of this process can be summed up in three possible scenarios:
\begin{enumerate}
	\item The inner horizon can tend to the origin asymptotically in $v$, either as an inverse polynomial, an inverse exponential or quicker (e.g. the exponential of an exponential discussed in \cite{Barcelo2022}). If this behaviour persists until curvature becomes Planckian, then this outcome can be considered effectively indistinguishable from the next.
	\item The inner horizon can plummet to the origin at a finite time $v$. This outcome cannot strictly be achieved solely with perturbations in the form of shells, but it has been argued to occur in the presence of a continuous matter distribution \cite{Brady1995,Page1991}. The result would be the formation of a spacelike singularity at finite $v$,  making the causal structure akin to the Schwarzschild geometry, but with a subsequently formed, weakly singular \cite{EllisSchmidt1977} Cauchy horizon.
	\item The inner horizon can stabilise at a finite radial position. This can occur if the central core has a stiffness which makes the inner horizon move less than what eq.~\eqref{ingredient1} requires. Then, classical perturbations alone would cease to be enough to destabilise this horizon.
\end{enumerate}

\subsection{Semiclassical inner horizon evolution}

As an alternative approach, one may choose to ignore classical perturbations around the inner horizon, and instead focus on its possible dynamics stemming solely from semiclassical backreaction. This approach may be justified if classical perturbations turn out to be negligible in comparison, and this indeed turns out to be the case in asymptotic analyses of analytical extensions of geometries with an inner horizon (in which the inner and Cauchy horizons coincide). These reveal that semiclassical backreaction in the vicinity of the Cauchy horizon may lead to corrections which are orders of magnitude larger than those stemming from classical mass inflation~\cite{Hollands2020a,Zilberman2022}. In other words, the singular nature of the RSET when approaching this horizon is stronger than that of classical matter, even when the latter has been amplified by the highly non-linear mass inflation effect. Although these analyses are not focused on dynamically formed inner horizons at finite times, they do provide some indication of what the dominant contribution to backreaction at late times (close to the Cauchy horizon) might be even in these scenarios.

A first analysis of the semiclassical evolution of the inner horizon at finite times was performed by the present author and collaborators in ref.~\cite{Barcelo2021}. There, we found that the RSET contains a negative ingoing energy flux, akin to the asymptotic $\expval{T_{vv}}$ term found in e.g. \cite{Hollands2020a}, related to the surface gravity of the inner horizon. Although as a source of dynamics in the semiclassical Einstein equation \eqref{semicl} this term is suppressed by a Planck constant, it is constant (in Eddington-Finkelstein coordinates) throughout the lifetime of a static inner horizon. When contrasted with the expected asymptotic decay in time of the classical ingoing perturbation, it is not surprising that the RSET is expected to produce a more drastic backreaction effect at late times.

This negative flux term can be interpreted in direct analogy to the negative flux driving Hawking evaporation at the outer horizon \cite{DFU}. Much like how Hawking's calculation depends on the exponential peeling off from the outer horizon of null trajectories, our result is directly related to the time-reversed of this behaviour at the inner horizon, i.e. the exponential accumulation of null rays, and its effect on the modes of a quantum field.

Backreaction from the Hawking effects leads to the inward displacement of the outer horizon, allowed due to the violation of energy positivity conditions in the RSET. Likewise, we found that backreaction at the inner horizon leads to a classically forbidden evolution: an outward displacement of this horizon, which shrinks the trapped region from the inside out. Additionally, unlike the slow Hawking evaporation of the outer horizon, the effect at the inner horizon seems to have a far more explosive accumulative character, likely due to the generally unstable nature of this horizon. The initial outward tendency of this horizon is exponential; hence, we have dubbed the effect \textit{inner horizon inflation}. If one extrapolates from this tendency, one finds that the trapped region would be extinguished in a time scale,
\begin{equation}
v_{\rm infl}\lesssim\frac{M}{M_\odot}\times 10^{-5}\,\text{s},
\end{equation}
with $M$ being the initial BH mass and $M_\odot$ the solar mass. 
This time is very short when compared with the extremely long Hawking evaporation time.
In fact, this behaviour suggests that the trapped region might be relevant only in the description of a transient phase in the path towards the formation of stationary ultracompact configurations without strict horizons~\cite{Visser2008,Carballo-Rubio2017,Arrechea2021}.

\subsection{Semiclassical effects over a mass inflation regime}

The central point of this essay is to illustrate how one may begin to provide a more holistic view on the structure of BHs in semiclassical gravity. The main example I present consists in the calculation of the RSET on a classical mass inflation background, and an analysis of the resulting competition between the mass inflation effect (which pushes the inner horizon inward) and semiclassical backreaction (which, as we have found, tends to push it outward).
Using a simple classical background which reproduces the causal behaviour of a mass inflation geometry, in our recent work~\cite{Barcelo2022} we tackled this very problem. Remarkably, we found that the negative ingoing flux related to the inner horizon indeed tends to push this horizon outward even in this dynamical background. 
More precisely, given that the semiclassical dynamics is initially suppressed by the Planck constant, classical mass inflation (when it occurs) can still dominate the initial behaviour of the system. However, semiclassical effects tend to become dominant at a later stage, seemingly taking control of the final fate of the inner horizon.

\begin{figure}
	\centering
	\includegraphics[scale=1]{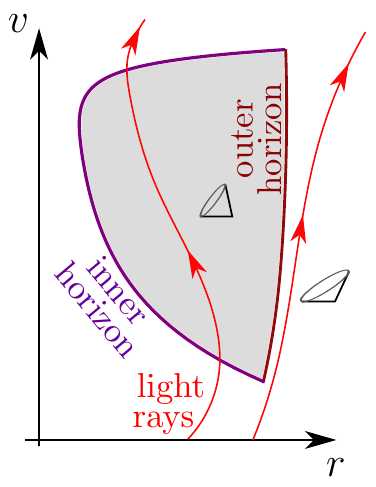}
	\caption{Evolution of a BH in which the inner horizon begins an inward motion due to classical mass inflation, but is then pushed outward due to semiclassical backreaction, eliminating the trapped region from the inside-out. The axes represent Eddington-Finkelstein coordinates.}
	\label{f2}
\end{figure}



Interestingly, the perturbation to the geometry we considered takes a form similar to the Reissner-Nordström mass term, but with the opposite sign, and growing much more quickly than its classical counterpart. It must be said, however, that the approximations we used in that model are only accurate up to the point at which semiclassical backreaction reaches the order of magnitude of the background terms (i.e. up to a point at which the classical inward movement of the inner horizon is nearly halted), but the result is indeed very suggestive of another manifestation of the inner horizon inflation we found in the static background case. This leads us to speculate that the trapped region may indeed be extinguished from the inside-out, as shown in fig.~\ref{f2}, instead of going through the slow Hawking evaporation from the outside.

\section{Towards a more holistic view}

The possibility of an inner-horizon inflation mechanism has already been suggested in past works~\cite{BalbinotPoisson93,Ori2019,Zilberman2022}, in terms of a growth of the radial coordinate on a Cauchy horizon in analytical extensions of BH spacetimes. However, these works did not elaborate on the final consequences this process might have, particularly regarding the potential complete elimination of the trapped region.

The example presented here provides a first glimpse into what semiclassical BH dynamics may look like when all aspects of the evolution of these objects are considered. If nothing else, it emphasises the importance of understanding well the admixture of classical and semiclassical effects, i.e. how the complex behaviour at the core of BHs combines with semiclassical backreaction to reveal the possibilities for the global structure of the spacetimes of these objects. This is especially relevant now, given the recent promotion of BHs into the ranks of astrophysically observable objects, in both the electromagnetic and gravitational spectrum. Having a complete model of their evolution, and thus understanding their ultimate nature, has never before been of such importance.

\section*{Acknowledgements}

I would like to thank Carlos Barceló, Raúl Carballo-Rubio and Luis J. Garay, as well as Julio Arrechea and Gerardo García-Moreno, for many helpful discussions on this topic. Financial support was provided by the Spanish Government through the projects PID2020-118159GB-C43 and PID-2020-118159GB-C44, and the fellowship FPU17/04471.




\nocite{*}
\bibliography{Bibliografia}
\bibliographystyle{ieeetr}

\end{document}